\documentstyle[aasms4,psfig]{article}

\def\gs{\mathrel{\raise0.35ex\hbox{$\scriptstyle >$}\kern-0.6em 
\lower0.40ex\hbox{{$\scriptstyle \sim$}}}}
\def\ls{\mathrel{\raise0.35ex\hbox{$\scriptstyle <$}\kern-0.6em 
\lower0.40ex\hbox{{$\scriptstyle \sim$}}}}
\def\et{\hbox{et al.}$\,$}
\received{---}
\revised{---}
\accepted{---}

\begin{document}
\title{The Mass/Light Ratio of Early-Type Galaxies: \\ Constraints from
Gravitational Lensing in the Rich Cluster AC114\footnotemark}

\footnotetext{Based on observations with the NASA/ESA {\it Hubble
Space Telescope} obtained at the Space Telescope Science Institute,
which is operated by the Association of Universities for Research in
stronomy Inc., under NASA contract NAS 5-26555.} 
              
\author{Priyamvada Natarajan\altaffilmark{1},
Jean-Paul Kneib\altaffilmark{2}, Ian Smail\altaffilmark{3} \& 
Richard S.\ Ellis\altaffilmark{1}}
\affil{\tiny 1) Institute of Astronomy, Madingley Road, Cambridge, CB3 0HA, UK}
\affil{\tiny 2) Observatoire Midi-Pyrenees, 14 Av.\ E.~Belin, 31400 Toulouse, France}
\affil{\tiny 3) Department of Physics, University of Durham, South Rd Durham, UK}

\begin{abstract}
We present a new wide field image of the distant cluster AC114
($z=0.31$) obtained with {\it Wide Field and Planetary Camera II}
onboard the {\it Hubble Space Telescope}. This image considerably
extends our knowledge of the lensing properties of the cluster beyond
that derived by Smail \et (1995a) from a single WF/PC-1 pointing. In
conjunction with published ground-based spectroscopy, we utilise
several newly-discovered multiple images to construct an improved mass
model for the central regions of the cluster. Using this model, we
apply the methodology introduced by Natarajan \& Kneib (1997) to
interpret local perturbations to the cluster shear field on small
scales resulting from mass associated with individual cluster
galaxies.  We use the lensing signal to place new constraints on the
average mass-to-light ratio and spatial extents of the dark matter
halos associated with morphologically-classified early-type cluster
members.  We find that the total mass of a fiducial $L^\ast$ cluster
spheroidal galaxy is largely contained within $\sim$ 15 kpc radius
halo ($\sim$~8--10 $R_e$) with a mass-to-light ratio
${M/L_V}\,\sim\,{15^{+10}_{-4}} $ (90 \% c.l.) in solar units within
this radius.  Comparisons with similar estimates for field galaxies
suggests that the cluster galaxies in AC114 may possess less extensive
and less massive halos. Additionally, we find some indication that, at
a fixed luminosity, S0 galaxies are less extended than ellipticals, 
suggesting a difference in the efficiency of tidal
stripping of different galaxy types.  We discuss the consequences of
our results in the context of models for the dynamical evolution of
cluster galaxies and the observational prospects for extending such
analyses.
\end{abstract}  


\section{Introduction}

The distortions induced in the shapes of faint background galaxies by
the gravitational lensing of a foreground cluster offers a powerful way of
reconstructing the mass distribution of the intervening lens (Mellier
\et 1997). While weak lensing studies have addressed the distribution
of mass on large scales ($\gs 300 $ kpc\footnote{Throughout the analysis we have assumed $\Omega = 1$;
$\Lambda = 0$; $H_0 = 50 $~km/s/Mpc which corresponds to an angular
scale of $1'' = 5.605$ kpc at $z=0.31$.}, Smail \et 1995b; Squires
\et 1996), analysis of strongly lensed features has allowed more
detailed modelling of the morphology of the total mass on small scales
($\gs 20 $ kpc) in the central regions of a few clusters, e.g.\
Cl0024+16 (Kassiola, Kovner \& Fort 1993) and A2218 (Kneib \et 1996).
These latter studies have suggested that mass concentrations associated
with individual cluster galaxies are needed to fit the detailed
geometry of some giant arcs.

This paper is concerned with examining the role of galaxy halos in
defining the distribution of mass in clusters on intermediate scales
($\simeq 50$ kpc) across a range of environments within 
clusters from the core regions to the lower density outskirts. Our 
technique considers perturbations associated with an ensemble of 
cluster galaxies within a smooth global cluster potential. Such perturbations
can be derived from the shape of the gravitational shear field estimated  
via the distorted forms of faint background galaxies. A strong motivation
for this work is the need to examine whether the mass-to-light ratio
($M/L$) of galaxies (measured within a large effective aperture) varies
significantly between high density cluster regions and the field. Such
an environmental variation of the $M/L$ ratio of a galaxy might be
expected if these galaxies presently found in dense regions suffered
more complex interaction histories leading to a redistribution of the
associated gaseous, stellar and dark matter components  (c.f.\ Moore et
al.\ 1996). One possibility is that the extended dark halo would be
preferentially removed and redistributed, leading to a reduction in the
$M/L$ ratio compared to that found for isolated galaxies of the same
morphological type.  However, the scale on which this redistribution
occurs (and hence the `granularity' of the resultant dark matter
distribution within the cluster) is unclear and has important
implications for our understanding of how clusters assemble and evolve.

Previous estimates of the global $M/L$ of galaxies on intermediate
scales relate primarily to field galaxies and have been obtained from
dynamical studies of their satellite systems (Zaritsky \et 1993), the
rotational motion of the HI gas (Sackett 1995), as well as the
kinematics of the stellar component (Rix \et 1997).  Recently weak
lensing analyses have been applied to provide constraints on the mass
distributions on intermediate-large scales around ensembles of field
galaxies (Brainerd, Blandford \& Smail 1996, hereafter BBS), a similar
approach to that used here. Their statistical analysis of the
alignment of the faint field population around bright foreground
galaxies gives a characteristic lensing mass of $M =
(2.0^{+2.4}_{-1.0}) \times 10^{12} M_\odot$ (90\% c.l.) within a
radius of 200 kpc (at a median galaxy redshift of $z\sim 0.5$). This
corresponds to a $M/L_V = 100^{+110}_{-60}$ in solar units
within their large
aperture.  Their analysis also provided weak limits on the scale of
the halo around these galaxies, a typical extent less than 20 kpc was
rejected at the 2 $\sigma$ level, while the upper limit on the halo
size was unconstrained, $r_t \gs 200 $ kpc. Modelling field galaxies
as truncated isothermal spheres, Dell'Antonio \& Tyson (1996; DT
hereafter) have obtained modest limits on the central velocity
dispersion and outer radius via a galaxy-galaxy lensing analysis of
the Hubble Deep Field. Using a very small aperture they
find a central velocity dispersion
$\sigma\,=\,185^{+30}_{-35}$ km s$^{-1}$, $r_{\rm outer}\,\geq\,30$
kpc, and an average mass enclosed within 20 kpc of
$(1.2_{+5.0}^{-5.4})\,\times\,{10^{12}}\,M_\odot$.

In this paper we examine whether it is observationally feasible to
detect the granularity in the cluster mass distribution and hence place
limits on the sizes and masses of dark matter halos associated with
typical galaxies {\bf within} a rich cluster. To do this we analyse a
new wide-field {\it Hubble Space Telescope} ({\it HST}) image of the rich cluster
AC114 ($z=0.31$) in \S 2. We construct a detailed model of the
large-scale mass distribution within the cluster using the numerous 
strongly-lensed features visible in the {\it HST} data in \S 3. The new 
image allows us to improve upon the earlier model presented by 
Smail \et (1995a) on the basis of pre-refurbishment {\it HST} data.
Of considerable importance
is the simplicity of our overall mass distribution in the central
regions; only a single, regular mass concentration is visible both in 
the lensing shear field or the cluster X-ray emission (Smail \et
1995a), making AC114 an ideal cluster for our purposes. 

We use techniques developed in the theoretical discussion given by 
Natarajan \& Kneib (1995, 1997; hereafter NK) and examining the 
weakly lensed arclets seen through the cluster we search for local 
perturbations in the cluster potential and correlate these with 
cluster galaxies of known morphology and luminosity. In this way 
we provide new constraints on the $M/L$ and extent of their dark halos. 
Section~4 describes the formalism used to place limits on the mass associated
with individual cluster galaxies, while \S 5 presents our results and
\S 6 discusses these and gives our main conclusions.

\section{Data}

The rich cluster AC114 ($z$=0.31) was observed with the {\it WFPC2} during 
Cycle 5 for a total of 24 orbits through the F702W filter. The observations
were mosaiced across four different pointings each of 16.8 ks total 
integration to give a contiguous field of $\sim 6.5' \times 3.9'$ (c.f.\ Fig.~1). 
In addition, by positioning one of the WFC chips on exactly the same 
region in two of the pointings, a deeper exposure (33.6 ks) was achieved 
for a $80'' \times 80''$ field in the cluster center. The six orbits at 
each pointing were split into 3 pairs, each of which was spatially offset 
by 3 WFC pixels to enable the removal of hot pixels and similar artefacts.

The data was reduced using the standard STScI pipeline followed by
alignment with integer pixel shifts and stacking based on the
the {\sc STSDAS} task {\sc CCREJ}. Cataloging and photometry of
objects on these coadded frames was accomplished using the Sextractor
image analysis package (Bertin \& Arnouts 1996). Image detection
proceeded by first convolving the data with a 0.3\,arcsec diameter box
filter and applying a surface brightness threshold of $\mu_{702}=25.0$ 
mag arcsec$^{-2}$ (corresponding to $1.3\sigma$ of the sky noise) 
and a minimum object area of 12 contiguous pixels. The Sextractor package
provides `Kron' magnitudes for each object which we have placed on the 
$R_{702}$ system using zero-points published by Holtzman \et (1995).

Unfortunately, due to an error in the scheduling of the observations,
the {\it HST} images have $\sim 4$--5$\times$ the nominal sky
background, restricting the limiting magnitude at which reliable image
shapes can be determined to $R_{702}=26$ rather than the expected
$R_{702} \sim 26.5$--27 (the {\it WFPC2} observations of the central
field will be repeated in Director's discretionary time in the near
future).  Although the surface density of useable background images
overall is considerably reduced, the gain over the earlier WF/PC-1
data is still substantial, particularly in the double pointing of the
center region.

We show in Fig.~1 the central area of the {\it HST} field,
as in A2218 (Kneib
\et 1996) the high resolution imaging provided by {\it HST} uncovers a wide
range of strongly lensed features in the center of the cluster lens. 
In Fig.~2 we provide more detailed views
of these features to highlight their lensed nature. 
The
most prominent feature, originally discovered by Smail \et (1995a) as
S1/S2, is the hook-shaped $z=1.86$ galaxy for which we can now locate
the missing third image S3 (Fig.~2). The geometrical configuration of
these three images, together with the spectroscopic redshift, provides
a reliable absolute calibration of the mass in the central regions of
the cluster (see \S 3). Based on this calibration and assuming the
general morphology of the cluster potential as indicated by the X-ray
map (Smail \et 1995a), we can derive estimates of the likely redshifts
of the other multiply-imaged sources in the field.  A major
breakthrough provided by the new {\it HST} image is the identification of
three 5-fold multiply-imaged systems A1--5 (with a predicted redshift
of $z=1.67\pm 0.15$ from our lens model), B1--5 ($z=1.17\pm 0.10$) and
D1--5 ($z=1.18\pm 0.10$).\footnote{As some of these
images were previously unmatched with their counterparts, it is
convenient to introduce a new nomenclature. We note that B2=A1; C2=A2;
C1=A4; A2=A5; B3=A6 where we give our nomenclature first and then 
Smail et al.'s.} The former includes a radial feature (A4--A5) which
provides an important constraint on the inner-most regions of the
cluster potential. B1--5 represents a three-image `cusp' configuration
perturbed by a nearby cluster elliptical, adding two more images
(B3--B4).  D1--5 is a cusp arc (3 images) perturbed by the two big
ellipticals near the cD galaxy.  Each perturbing galaxy adds 2 images
to the system, but only one is visible, as the other one is
de-amplified and lies at the center of each galaxy.  Arguably the most
startling multiply-imaged feature in the cluster is C1-3, three
images of a partial-ring shaped galaxy which we estimate to lie at
$z=2.1\pm 0.3$.  The highly elongated morphology of the source, with
several bright knots lying along the ridge, provides an unique insight
into the morphological nature of galaxy formation at high redshift. A
number of other fainter candidate multiple images are visible around
the cD. We postpone detailed discussion of these features until the
deeper pointing of this field has been acquired.
We demonstrate
below that the lensed features catalogued here (Fig.~1, 2) are
sufficient to allow us to both improve the mass model over the earlier
one developed by Smail et al.\ (1995a) and to search for the
granularity due to individual cluster galaxies on a statistical basis.

The image catalog constructed from our frames contains $2446$ objects
brighter than $R_{702}=26$ over the 25 sq.\ arcmin of the field.  We
have adopted a bright magnitude cut-off of $R_{702}=23$ for our
background field sample in order to reduce contamination by foreground
and cluster galaxies.  Within $23\,<\,R_{702}\,<\,26$, all $1762$
galaxies were therefore considered as background. The $R_{702}$
magnitudes were converted into the rest-frame V-band assuming a
non-evolving E/S0 spectral energy distribution to determine the
k-correction. To select perturbing galaxies within the cluster we made
use of the morphologically-classified catalogue of Couch \et
(1997). To study the morphological mix within the cluster these
workers determined visual classifications for 485 galaxies brighter
than $R_{702} = 23.0$ ($M_V = -16.3$, assuming a non-evolving
elliptical spectral energy distribution) in the field of AC114. A
subset of 208 were classified as spheroidal (E or S0). In detail, the
morphological breakdown to $R_{702}$=23.0 is as follows: E (52), E/S0
(69) and S0+S0/a (87) (Couch et al.\ 1997). The median luminosity for
the sample limited at this apparent magnitude is close to the
characteristic luminosity $L^{*}$ of the cluster population. To a
limit of $R_{702} = 21.0$ the equivalent numbers are: E (22), E/S0
(10) and S0+S0/a (44). The central cD was removed from further
consideration since its halo is expected to differ significantly from
that of the average cluster member.

\section{Cluster Mass Modelling}

Considerable progress has been made in the interpretation of the
observed shear field in clusters, using the inversion technique first
described by Kaiser \& Squires (1993), a refined version developed by
Seitz \& Schneider (1995) as well as through local finite field
inversion techniques suggested by Kaiser (1995) and Schneider
(1995). These methods are appropriate for low resolution mapping in
the weak regime. Better resolution is possible in the central regions
of the cluster where strong lensing effects are prominent.  Our
approach following NK (described below) extends both the $\chi^2$
model fitting of Kneib \et (1996) as well as building upon the
techniques of BBS and their maximum likelihood extension (Schneider \&
Rix 1997).  Both strong and weak lensing constraints are included in
this mass reconstruction procedure for clusters as traced by the
observed shear field.

An accurate mass model for AC114 is an essential prerequisite to the
various galaxy-galaxy lensing methods discussed in $\S$4.  The large
number of multiply-imaged features identified from the new {\it WFPC2}
imaging of AC114 means that an accurate model is feasible, especially
when these multiply-imaged features are combined with the observed
radial profile of the smoothed tangential shear. Here we follow
closely the methodology of Kneib \et (1996), constraining the
morphology of the cluster mass distribution with the positions and
relative amplifications of the 5 sets of multiple images (above) as
well as the large-scale weak shear field.  The iso-density contours
are shown in Fig.~1. The simplicity of the cluster mass distribution
as indicated by both the lensing shear field and the cluster X-ray
emission (Smail \et 1995a), make AC114 an ideal cluster for our
purposes. Further detailed comparison of our lensing model with
numerical cluster simulations (including the hydro-dynamics of the
gas) as well as the X-ray data will be dealt with in a forthcoming
paper. Our best-fit model for the central regions ($r\,\leq\,250\,$
kpc) has a single, dominant potential centered on the cD, with an
ellipticity and orientation close to that of the cD halo (Table~1).
The total masses within 75; 150; 500 kpc of the cluster centre are
respectively $(0.42\pm{0.01}) \times 10^{14} M_\odot$ ; $(1.2\pm
{0.15}) \times 10^{14} M_\odot$ and $(4.0\pm {0.4}) \times 10^{14}
M_\odot$. The total mass-to-light ratio for AC114 out to 500 kpc is
found to be ${M/L_V}\,\sim\,{160} (M/L_V)_\odot$.

\section{Galaxy-Galaxy Lensing Methods}

\subsection{Formalism}

The cluster potential ($\phi_{\rm tot}$) in AC114 is treated as a
linear superposition of a smooth large-scale potential ($\phi_{\rm c}$)
with a typical scale $>$ 20 arcsec and several small-scale
perturbations ($\phi_{\rm cgal_i}$) representing cluster
galaxies and their associated halos, i.e.\  
\begin{eqnarray} \phi_{\rm
tot} = \phi_{\rm c} + \Sigma_i \,\phi_{\rm cgal_i}, \label{eq1}
\end{eqnarray} 
As shown by NK, the complex shear $g$ evaluated for such
a potential in the frame of the perturber can be written as a sum of
contributions arising from the intrinsic shapes of the sources, that
induced by the smooth cluster component and that from the small-scale
perturbers (see $\S$2.2 of NK for details). The small-scale potentials
are represented by pseudo-isothermal elliptical mass distributions
(Kassiola \& Kovner 1993) and each is characterised by a central
velocity dispersion ($\sigma_0$), a core-radius ($r_0$) set to be 
0.15 kpc for a $L^\ast$ galaxy and a
truncation radius ($r_t$). To minimize the number of parameters needed
to characterize a fiducial galaxy, a set of scaling laws physically 
motivated by the Fundamental Plane (FP), similar to those used by BBS 
are adopted:
\begin{eqnarray}
{\sigma_0}\,=\,{\sigma_{0*}}({L \over L^\ast})^{1 \over 4};\,\,\,
{r_0}\,=\,{r_{0*}}{({L \over L^\ast}) ^{1 \over 2}};\,\,\,
{r_t}\,=\,{r_{t*}}{({L \over L^\ast})^{\alpha}}.
\label{eq2}
\end{eqnarray}
The total mass, $M$, and the mass-to-light ratio, $\Upsilon$, scale as:
\begin{eqnarray}
\,\,M\,=\,{2 \pi {\Sigma_0} {r_0} {r_t}}\,= 
{9 \over {2 G}}{{\sigma_{0*}}^2}{r_{t*}}({L \over L^\ast})^{{\frac 1 2} + \alpha}\,;
\label{eq3}
\end{eqnarray}
\begin{eqnarray}
{\Upsilon}\,= 21\, \left( {\sigma_{0*}\over 240\,{\rm km\,s^{-1}}}\right)^2
   \left( {r_{t*}\over 15\,{\rm kpc}} \right)\,\left({\frac
   {L}{L^\ast}} \right)^{\alpha - {\frac 1 2}},
\label{eq4}
\end{eqnarray}
where $\Sigma_0$ represents the projected mass density at the lens and
$\alpha$ is an index that is assigned a value of 0.5 yielding the constant
mass-to-light ratio case or a value of 0.8 leading to
$\Upsilon\,\sim\,{L^{0.3}}$ consistent with the observed correlations on
the FP (e.g.\ Jorgensen \et 1996).

\subsection{Direct Aperture Averaging Method}

The simplest approach toward measuring the granularity in the shear pattern
is to average the mean gravitational shear within apertures around each
of the cluster galaxies. This `direct' method does not require an
accurate description of the smooth cluster mass distribution and
consequently yields only approximate limits on the mean $M/L$ ratio on
intermediate scales. Nevertheless, in the weak regime, the large scale
shear contribution will vary at most linearly across a given annulus,
making subtraction straightforward. The residual tangential component
of the shear in the frame of an individual perturbing galaxy then
constrains the typical galaxy $M/L$ through the various scaling
relationships given in $\S$3.1. The signal to noise of these
measurements depends not only upon the masses of the galaxies, but also
on a number of other factors of which the surface density of background
sources is the most important (NK). 

\subsection{Maximum Likelihood Method}

While the direct averaging method is straightforward to apply, to
robustly and independently constrain both of the fiducial parameters of
interest ($\sigma_{0*}$ and $r_t$) the large-scale mass distribution
needs to be taken into account in the non-linear, over-critical
central regions of the cluster. The maximum likelihood method proposed in
NK allows us to do this and we now describe its application to our
observations of AC114. The essential procedure is to maximize the
likelihood function of the predicted distribution for the ellipticities
of the background sources, using a set of model parameters for the
cluster galaxies, given the distributions of intrinsic ellipticity and
redshift for the faint field population.  In the weak regime, the
intrinsic ellipticity vector $\tau_{S_j}$ of the $j$th faint background
galaxy can be determined from its observed ellipticity vector
$\tau_{\rm obs_j}$ by subtracting the induced shear vector from the
various components of the cluster lens:
\begin{eqnarray}
\tau_{S_j} \,=\,\tau_{\rm obs_j}\,-{\Sigma_i^{N_{\rm
      cgal}}}\,{\gamma_{{\rm cgal}_i}}\,-\,
 \gamma_{\rm c},
\label{eq5}
\end{eqnarray}
where $\Sigma_{i}^{N_{\rm cgal}}\,{\gamma_{{\rm cgal}_i}}$ is the sum
of the shear contribution at a given position $j$ from $N_{\rm cgal}$
perturbers, and the term $\gamma_{\rm c}$ is the shear induced by the
smooth cluster component. Note here that the expression in eqn. (5)
above is the form obtained in the limiting case of the weak regime - 
where the effects of magnification are small, for the maximum 
likelihood analysis, we use the full non-linear relation to incorporate
both the strong and weak lensing regimes as described in NK. For each
arclet, a redshift is assigned from the expected redshift distribution
$N(z)$ corresponding to its intrinsic magnitude. Limits on the $N(z)$
at such faint limits can be provided by classical spectroscopy (Cowie
\et 1996), with alternative estimates provided through inversion
techniques employing highly-constrained cluster lenses
(Kneib \et 1996, 1997), and photometric modelling of the Hubble Deep
Field (e.g.\ Metcalfe \et 1996; Mobasher \et 1996; Lanzetta \et 1996;
Connolly \et 1997).  We adopt the $N(z)$ derived for an $R_{702}\sim
25$ distribution by Kneib \et (1996) extrapolating it slightly using
the no evolution predictions when necessary. For moderate redshift
cluster lenses the uncertainties associated with the redshift
distribution of background galaxies are no longer a major concern
(Ellis 1997). The intrinsic shape for the background source is also
randomly assigned from the distribution measured in blank fields
imaged with {\it WFPC2} using the Medium Deep Survey data and other
similar {\it HST} archival {\it WFPC2} data (Ebbels \et 1998).

Given a detailed mass model for the cluster, $\gamma_{\rm c}$ can
be specified so the likelihood for a proposed self-similar parameter
set for the cluster galaxies (eqn.~2) can be expressed as:
\begin{eqnarray}
 {\cal L}({{\sigma_{0*}}},{r_{t*}}) = 
\Pi_j^{N_{bgal}} p(\tau_{S_j}).
\label{eq6}
\end{eqnarray}
The likelihood function of the estimated probability distribution of
the source ellipticities (given that the functional form of the
intrinsic ellipticity distribution is assumed to be known for field
galaxies from the MDS) is maximized with respect to the model  
parameters (${\sigma_{0*}},{r_{t*}}$) and is optimized for agreement
with the constraints from the strong lensing (the sets of multiple images)
in the central regions. 

In principle, the likelihood function ${\cal L}$ needs to be computed
for many different realizations, i.e. the redshift drawn for any
individual image in fact needs to be the mean over different draws.
However, we demonstrate below that it is equivalent to constructing
the likelihood for a single realization where the redshift ${z_{j}}$
of the arclet drawn is the median redshift corresponding to the
observed source magnitude from $N(z,m_j)$.  On performing a
Monte-Carlo sum over $N_{\rm MC}$ realizations of $z_j$, the
likelihood becomes:
\begin{eqnarray}
{\cal L}({{\sigma_{0*}}},{r_{t*}}, ...) = 
\Pi_j^{N_{gal}} {\Pi_k^{N_{\rm MC}}}p(\tau_{S_j^k}),
\end{eqnarray} 
where $p_{\tau}(\tau_{S_j^k})$ is the probability of the source
ellipticity distribution at the position $j$ for $k$ drawings
for the redshift of an arclet of magnitude $m_j$. The mean over
 $N_{\rm MC}$ realizations gives:
\begin{eqnarray}
\left<{p(\tau_{S_j})}\right>\,=\,{1 \over {N_{\rm
MC}}}\,{\Sigma_{k=1}^{N_{\rm MC}}}\,{p(\tau_{S_j^k})}.
\end{eqnarray}
It is equivalent to the following expression in integral form,
\begin{eqnarray}
\left<{p(\tau_{S_j})}\right>\,&=&\,{{\int {{p(\tau_{S_j}(z))}{N(z,{m_j})\,dz}}} \over
{\int {N(z,{m_j})\,dz}}}\,\\ \nonumber &=&\,{p(\tau_{S_j}({z_{\rm avg}}))}\,\sim\,{p(\tau_{S_j}({z_{\rm median}}))}
\end{eqnarray} 
${z_{\rm avg}}$ being the average redshift corresponding to the
magnitude $m_j$. We define ${z_{\rm avg}}$ so that
${p(\tau_{S_j}({z_{\rm avg}}))}$ is equal to
$\left<{p(\tau_{S_j})}\right>$, and since both the ellipticity
distribution and the redshift distribution are `broad' ${z_{\rm avg}}$
is nearly equal to ${z_{\rm median}}$ - the median redshift of the
magnitude limited arclet sample. Therefore the corresponding
likelihood ${\cal L}$ can then be simply written as,
\begin{eqnarray}
{\cal L}\,=\,{\Pi_j}{\left<{p({\tau_{S_j}})}\right>}
\end{eqnarray}
and the log-likelihood as $l\,=\,\ln {\cal
L}\,=\,{\Sigma}{\left<{p(\tau_{S_j})}\right>}$. The best estimate of
the model parameters are obtained by maximizing this log-likelihood
function with respect to ${\sigma_{0*}}$ and ${r_{t*}}$. Geiger \&
Schneider (1997) have also recently demonstrated the robustness of 
such a maximum-likelihood implementation.

~From the simulations presented in Section~5.4 of NK, it was found that
for robust results the cluster model used to derive $\gamma_{\rm c}$
(based upon strongly-lensed features) must have a centre known to
$\pm$ 5 arcsec and an Einstein radius to within 20\%.  Our model
of AC114 discussed in \S3 attains the necessary accuracy (c.f.\ Table~1).

\section{Results}

We first present the results from our direct estimation of
galaxy-galaxy lensing in AC114.  Direct averaging of the shear field
on small scales around the 208 E and S0 cluster galaxies defined as
perturbers in our analysis does provide a weak detection. The signal
is computed in annuli centred around the perturbing galaxies.  The
strongest signal as expected lies within the innermost bin,
corresponding to a radius of less than 1.5 arcsec. The measured value
of the shear there is $\tau\,=\,{0.16}_{-0.13}^{+0.12}$ (1$\sigma$).
Simulating the composite mass distribution in AC114 (i.e.\ including
the cluster members and the model described in \S~3), using the
methodology described in \S~4 of NK, this measured value of $\tau$ is
translated into a $M/L_V$. For a $L^\ast$ galaxy the physical radius that
corresponds to the outer-most annulus is $\sim 35 $ kpc.  Subdividing
the sample of cluster galaxies into morphological classes, we find
that the signal is dominated by the brighter, spheroidal galaxies for
which we find $M/L_V\,\sim\,{16_{-12}^{+8}} (M/L_V)_\odot$ within 35
kpc.  These errors have been assigned on the basis of simulations of
the constructed composite mass model (main clump + cluster galaxies),
and therefore include the effects of the assumed linearity of the
potential induced by this smooth clump on large scales.

Turning to our maximum likelihood analysis, we also report a signal detection
(Figs~3). These figures show that we can set joint constraints on
both the central masses of the cluster galaxies and their halo sizes.
In Fig. 3a, we utilised the entire sample of cluster galaxies adopting
$\alpha = 0.5$ which yields a well-defined likelihood peak at
$\sigma_{0}^{*} = 192_{-27}^{+53}$ km s$^{-1}$ ($\sigma_{0}^{*}\ > 250$
km s$^{-1}$ is excluded at 95\% c.l.) and $r_t \approx 17 $ kpc. These
values translate into a total mass for an $L^\ast$ galaxy of $M\,\sim\,
{(4.9^{+3.1}_{-1.3})} \times {10^{11}} \,M_\odot$ and an average $M/L_V
\sim {15^{+10}_{-4}} (M/L)_\odot$.  Note that only the dispersion in the
value of $\sigma_{0}$ is used in the computation of the 90\% c.l. error
bars quoted above.  

In Fig.~3b, we illustrate the effect of adopting $\alpha$ = 0.8. In
this case, the likelihood has a maximum at $\sigma_{0}^{*} =
208_{-12}^{+17}$ km s$^{-1}$ ($\sigma_{0}^{*}\ > 250$ km s$^{-1}$ once
again excluded at 95\% c.l.) and $r_t \approx 18 $ kpc. These values
translate into a typical total mass for an ${L^\ast}$ galaxy of
$M\,\sim\,{(6.1_{+1.1}^{-0.7})} \times {10^{11}} \,M_{\odot}$ or an
average $M/L_V \sim {19_{-2}^{+4}} $ in solar units.  

Taking $\alpha=0.5$ and fixing the fiducial halo size obtained above
($r_t = 17 $ kpc), we finally split the sample into E's and S0's and
compute the likelihood estimate for their characteristic velocity
dispersions at a fixed luminosity independently. We note here that the
median luminosities and ranges for the two categories are very
similar. We find a marginal indication that the E galaxies are 
more massive than the S0s within our effective radius:
$\sigma_{0}(E)\,=\,{190_{-52}^{+15}}$ km s$^{-1}$ and
$\sigma_{0}(S0)\,=\,{120_{-32}^{+17}}$ km s$^{-1}$ (Fig.~3c). 
At the 60 \% c.l., E's and S0's seem to have differing characteristic
central velocity dispersions, which can be attributed to arising from
structural differences, for instance, due to differences in their 
truncation radii.

The reliability of these conclusions has been tested via extensive
simulations and tests. In particular we have conducted several null
tests by randomising (i) the orientation of the background sources,
(ii) the positions of the background sources within the frame, and
(iii) the lens centers. In each case no significant maxima were
spuriously produced in the likelihood test. The principal sources of
error in the $M/L_V$ estimates are (i) shot noise - we are inherently
limited by the finite number of sources sampled within a few tidal
radii of each lensing cluster galaxy, (ii) the spread in the intrinsic
ellipticity distribution of the source population, (iii) the unknown
source redshifts and (iv) observational errors arising from
uncertainities in the measurement of ellipticities from the images for
the faintest objects.

In their simulation NK (Section 4) showed that the strength of the
signal and hence the significance of detection increases with
increasing the mass-to-light ratio of the cluster galaxies as well as
increasing the total number of background galaxies. NK demonstrated
(Section 4.3) that a cluster/background number ratio $r$ of 0.02 would
ensure a 3-$\sigma$ detection for $\Upsilon \gs 4$ for a cluster at
z=0.175 and that the detection significance improved with
$\sqrt{N_{bg}}$ as expected. However, these simulations, whilst
realistic to the extent of incorporating the statistical properties of
foreground and background galaxies according to various simple models,
did not incorporate uncertainties arising from shot noise, possible
uncertainties in estimates of the redshift distribution of background
galaxies and instrumental effects such as the sampling and point
spread function characteristics of WFPC-2. The unknown redshift
distribution of the background sources will affect the retrieval of
the signal in a systematic way. For instance, the total mass estimate
for the cluster galaxies for a given shear would be under-estimated 
if the mean redshift of the background population was increased.

In the context of the HST data for AC114, we have therefore revisited
these simulations. For the actual cluster/background ratio adopted in
the observational analysis of AC114 ($r \sim$ 0.05 - since the bulk of
the signal arises predominantly from the brighter half of the
perturbers), running NK's simulations we find that noise-free
simulations would yield a 3-$\sigma$ detection for $\Upsilon = 15$.
In practice, the maximum likelihood contours in Fig.~3 yield a
confidence level of approximately 1.7-$\sigma$ illustrating the
possible degradation introduced by noise.

We have also separated our sample of perturbers on the basis of their
luminosity (bright/faint) and find that the majority of the power in
our tests is, unsurprisingly, coming from the more luminous half of
the sample, the mix of E vs S0 in the two samples is bright (36:40)
and faint (34:44). Separating the sample using the radial distance
from the cluster centre (inner/outer) showed no statistically
significant differences between the two samples, again the mix of E vs
S0 in the two samples is inner (32:37) and outer (49:37). The absence
of any measurable radial variation in the properties of the cluster
galaxy halos in our sample is not surprising given the limited sample
size and area coverage. On combining comparable data sets for each of
10 clusters observed with the forthcoming {\it HST} {\it ACS} ({\em Advanced
Camera for Survey}), our techniques would provide 99.5\% confidence
measurement of the fiducial parameters opening the way for the
extension of this analysis to study the radial variation of the
mass-to-light ratio within clusters.

\section{Discussion and Conclusions}

In this paper, we have reported the first detection of the signature
of extended dark halos around galaxies in the rich cluster
AC114 at $z=0.31$. ~From our analysis we find that a $L^\ast$ spheroidal
cluster galaxy has a total mass of $M \sim (4.9^{+3.1}_{-1.3}) \times
{10^{11}}\,M_\odot$, and a $M/L_V\sim {15^{+10}_{-4}} (M/L)_\odot$.
We find that the {\bf total} mass of a
fiducial $L^\ast$ cluster spheroidal is primarily contained with $\sim$
15 kpc, with some indication that the halos of the S0 population may be 
more truncated than those of the ellipticals.

For field galaxies, with greater uncertainty, galaxy-galaxy lensing
has been used to place constraints on halo sizes and masses of field
galaxies on larger scales (BBS \& DT). Comparing our results with
those from both these field studies, 
we find an apparent halo cut-off in the cluster which is
not seen in the field, $r_t\sim 15$ kpc in AC114 
versus $r_t\gs 200 $ kpc (BBS) and
$r_t\gs 30 $ kpc (DT) for the field.  This
would give some support to the hypothesis that the cluster
galaxies have truncated mass profiles, as expected if
they suffer substantial stripping of their dark matter halos during
the formation of the cluster.  Moreover, the evidence for
smaller truncation radii in the S0 galaxies compared to the ellipticals
would imply that they have suffered more stripping, possibly
due to their less bulge-dominated (compact) mass distributions.
These differences in the 
truncation radii found by the various studies lead to
differences in the total masses inferred inside large apertures (larger
than the truncation radii of some of the samples).
We caution, however, that differences in
the sample selection, observing conditions and methodologies for these
studies make any accurate
comparison complex. A similar {\it HST}-based survey of
weak lensing by elliptical galaxies in the field (Ebbels \et 1998)
should provide the optimal field sample for detailed comparison with
our results. The prospects for comparing the variation of the
mass-to-light ratio for galaxies as a function of environment are
promising.

Comparing the total mass in cluster galaxies within 250 kpc of the
center (down to the magnitude limit of our selection criterion) 
to the total mass of the cluster, we estimate that approximately 11\%
of the mass of the cluster is bound to individual cluster galaxies.
The fraction of the total cluster mass that is associated with individual
galaxies has important consequences for the rate of galaxy interactions
and hence the evolution of the cluster on the whole. The tidal field of
the cluster potential well is strong enough to truncate the dark matter
halo of a galaxy whose orbit penetrates the cluster core.  In the context
of this truncation picture, the tidally limited radius ($r_{\rm tidal}$)
and mass of a cluster galaxy whose orbital pericenter is close to the
cluster core radius can be estimated from (Merritt 1988):
\begin{eqnarray}
{r_{\rm tidal}}\,\approx\,\,30\,\,(\frac {\sigma_{0*}}{240{\rm \,km\,s}^{-1}})\,
({\frac {\rho_0}{0.04 \,M_\odot{\rm\,pc}^{-3}}})^{-\frac{1}{2}}\,\,{\rm kpc},
\end{eqnarray}
\begin{eqnarray}
M\,\approx\,{1.4\,\times\,{10^{12}}}\,{(\frac {\sigma_{0*}}{240{\rm \,km\,s}^{-1}})^3}\,
(\frac {\rho_0}{0.04 \,M_\odot\,{{\rm pc}^{-3}}})^{-\frac{1}{2}}\,M_\odot
\end{eqnarray}
where $\rho_0$ is the central density of the cluster.  Using the
estimate of $\rho_0$ for AC114 from the strong lensing model ($\rho_0$
= 0.0375 $M_\odot\,{pc}^{-3}$ and $\sigma_0$ from the
maximum-likelihood analysis, we obtain ${r_{\rm tidal}}\,\sim\, 25 $
kpc and an estimate of the mass enclosed within $r_{\rm tidal}$, $M
\,\sim\, {9.0 \times {10^{11}}} \,M_\odot$; which compares well with
the values obtained from the maximum likelihood analysis (${r_{\rm
t}}\,\sim\, 17 $ kpc and $M \,\sim\, {(4.9^{+3.1}_{-1.3}) \times
{10^{11}}} \,M_\odot$).  Dark halos of the scale of $\sim 15$--30 kpc
indicate a high probability of galaxy--galaxy collisions over a Hubble
time within a rich cluster. However, since the internal velocity
dispersions of these cluster galaxies ($\sim 150$--250 km s$^{-1}$)
are much smaller than their orbital velocities, these interactions are
unlikely to lead to mergers, suggesting that the encounters of the
kind simulated by Moore \et (1996) are the most frequent and likely.

These first results on the properties of galaxy halos within clusters
from lensing are very encouraging.  We are therefore extending our
analysis using both observations of galaxies in the central regions of
rich clusters at $z=0.17$--0.56 (Natarajan, Kneib \& Smail 1998) and
across a range of environments within a number of clusters at $z\sim
0.3$.  These two samples will provide insights into the role of
changes in halo properties in the evolution of both cluster spheroids
(Ellis \et 1997) and disk galaxies (Couch \et 1997), as well as the
variation of these effects with local environment.  The hope of future
expansion of this technique appears good with the proposed
installation of the the {\it ACS} in the next {\it HST} Servicing Mission, which
will be able to cover a wider field providing the ideal data-sets for
such studies.  Similarly, progress is expected in the techniques and
formalism used to constrain the mass distribution on different scales,
taking full account of the noise in the ellipticity measurements.
Therefore, the prospects for understanding the evolution of galaxy
halos from field galaxies to cluster cores and hence the radial
variation of the mass-to-light ratio are promising.

\section*{Acknowledgements}

We acknowledge useful discussions during various stages of this work
with Neta Bahcall, Tereasa Brainerd, Tim Ebbels, Marijn Franx, Josh
Frieman, Bernard Geiger, Nick Kaiser, Chris Kochanek, Martin Rees,
Peter Schneider and Simon White. JPK acknowledges support from the
CNRS and IRS acknowledges funding from the Particle Physics and Astronomy
Research Council.

\newpage
\begin{figure}
\caption{The central regions of our {\it WFPC2} image of AC114.  The
area displayed is roughly $560\times 560$ kpc in extent.  We identify
the various multiply-imaged features in the center of the cluster
(c.f.\ Fig.~2 and \S 2).  We also overplot the iso-density contours of
our best fitting mass model for the cluster constrained by the
geometry and redshifts of these features. The details of the model
mass distribution are given in Table~1. The overplotted shear field
corresponds to the model predictions for sources placed at $z = 1$.}
\end{figure}


\begin{figure}
\caption{A mosaic of the brighter
multiply-imaged sources seen in the central region of AC114 is presented
here.  Starting with S1/S2 (top), these represents two images of a
$z=1.86$ galaxy, identified and spectroscopically confirmed by Smail
\et (1995a). However, they were unable to identify the counter-image
S3 in their search of multi-colour ground-based images due to its
proximity to a bright cluster elliptical.  We show all three images of
this remarkable object with a compact core and blue hook-shaped
extension, which illustrates the striking symmetry shown by the
various images.  Next we present five images of a clumpy background
source: A1--5, A4/A5 representing two images of the source merging
across the radial critical line.  Our detailed mass model for the
cluster, normalised using the observed position of the critical line
for source `S' at $z=1.86$, indicates that the source `A' lies at
$z=1.67\pm 0.15$ and provides a probable core radius for the cluster
mass of $r_c \sim 50$ kpc.  For the B images we predict $z=1.17\pm
0.10$.  We note that the image B3/B4 corresponds to two merging images
of the source, and its apparent higher surface brightness arises
because of the unresolved nature of the source at the resolution of
{\it WFPC2}.  Finally, we come to the striking three image system,
C1/C2/C3, with an estimated source redshift of $z=2.1\pm 0.3$.  This
partial-ring galaxy shows considerable internal structure on scales
corresponding to $\sim 200$ pc at the source. Additionally, we find
another candidate multiple image system (5 images) D1--D5 with
predicted redshift $z=1.18 \pm 0.10$. The source of images B and D
have similar redshift and may have physical links.  In each case the
tick marks represent 1 arcsec.}
\end{figure}

\begin{figure}
\caption{Panel [a]: Maximum-likelihood retrieval of the fiducial
parameters for cluster galaxies in AC114: the likelihood peaks at
${r_t^*}\,\sim\,$ 20 kpc and ${\sigma_0^*} \,\sim\,$190 km s${^{-1}}$
(marked by the intersecting dotted lines); and the inner 3 contour
levels correspond respectively to 60\%, 80\% and 90\% confidence
limits The galaxy models used in this case correspond to constant
mass-to-light ratio (i.e.\ ${\alpha} = 0.5$ was assumed in the scaling
relations) which are the overplotted solid curves for $M/L_V$ = 10, 17
\& 23 (increasing from bottom to top). The secondary peak at
$r_t\,\sim\,$ 50 kpc is clearly insignificant (since the relevant
contour demarcates only a 60\% c. l. region), however if that higher
value of $r_t$ is to be taken seriously then it implies a mass that is
larger by a factor of 2; Panel [b]: The fiducial parameters for the
case when the the mass-to-light ratio scales with luminosity, i.e.\
$\alpha = 0.8$ in the assumed scaling relations. The recovered
parameters with the maximum-likelihood in this case are: $r_t^* \sim
20$ kpc and $\sigma_0^* \sim 210$ km s${^{-1}}$ (marked by the
intersecting dotted lines); and the inner 3 contour levels correspond
respectively to 60\%, 80\% and 90\% confidence limits and once again
the overplotted solid curves correspond to the loci of constant
$M/{L_V}$ = 13, 23 \& 27 (increasing from bottom to top) for a
$L^\ast$ galaxy; Panel [c]: Parameters recovered for the sub-samples:
the cluster galaxies were split into 2 primary morphological classes
the E's and the S0's. Combined optimization for their respective
fiducial velocity dispersions yields: ${\sigma_0^*}(E)\,\sim\,$ 190 km
s${^{-1}}$ and ${\sigma_0^*}(S0)\,\sim\,$ 120 km s${^{-1}}$. The inner
3 contours correspond to 60\%, 80\% and 90\% confidence limits. While
the secondary peak seen is probably induced by the noise, the contours
imply that at the 90\% c.l. level the masses (or equivalently the
truncation radii) of the Es and S0s are indistinguishable.}
\end{figure}

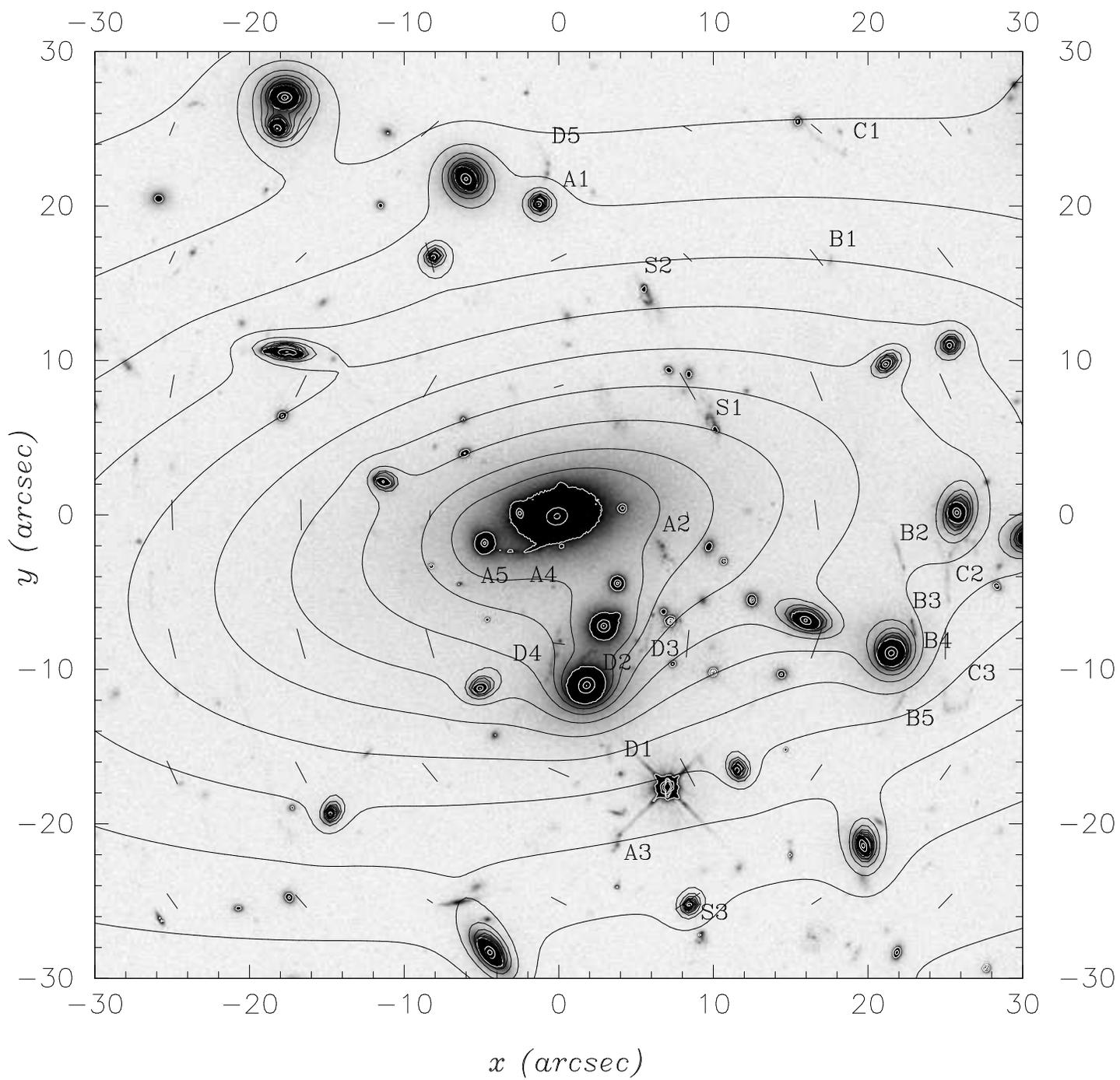
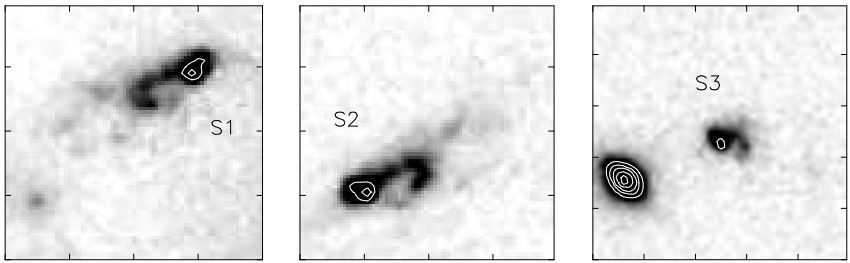
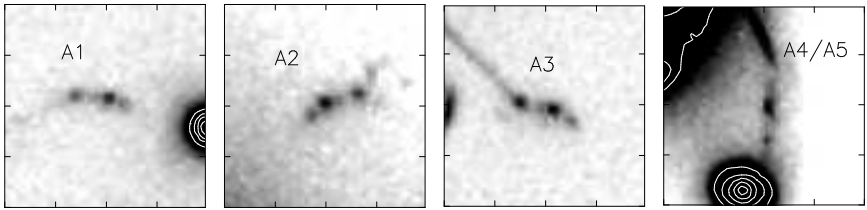
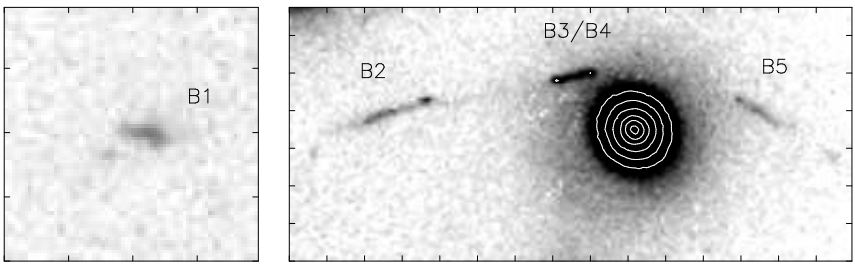
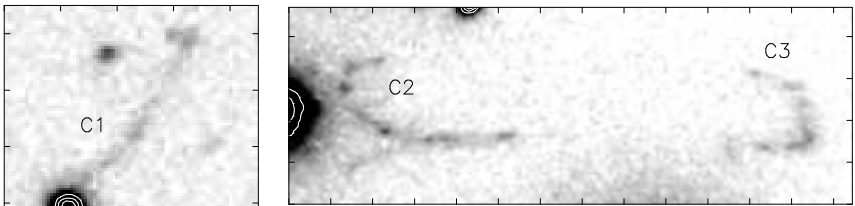
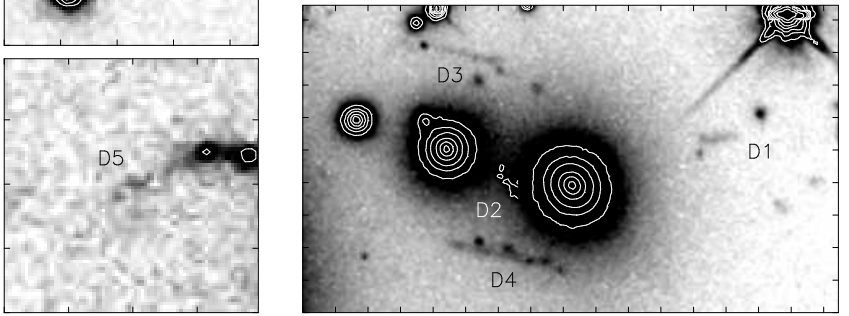
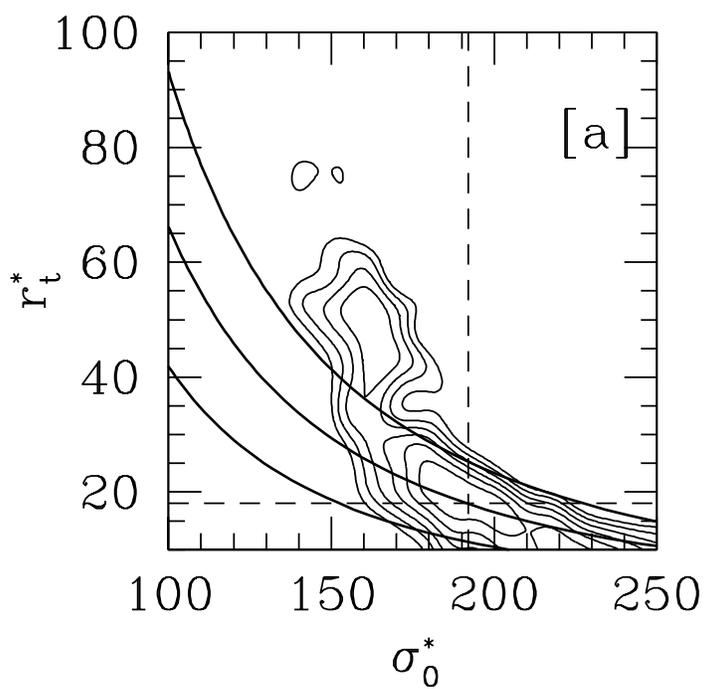
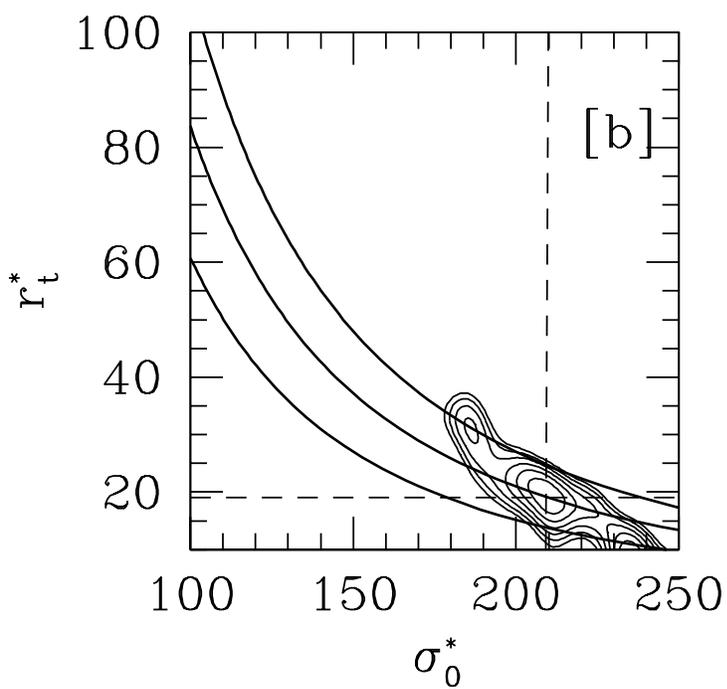
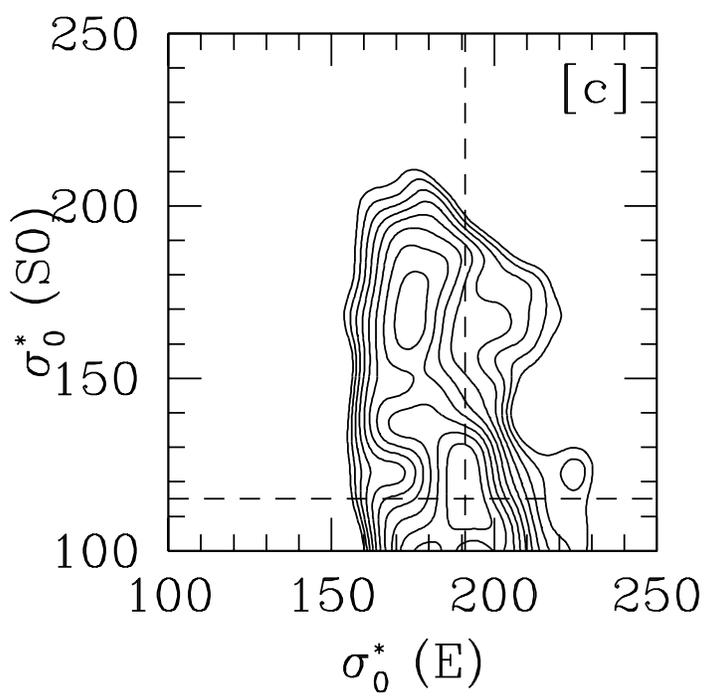
\end{document}